\documentclass[runningheads]{llncs}
\usepackage[utf8]{inputenc}
\usepackage{textcomp}
\usepackage{graphicx,verbatim}
\usepackage{hyperref}
\usepackage{cite}
\usepackage{amsmath}
\usepackage{amssymb}
\usepackage{authblk}
\usepackage{bbding}
\usepackage{graphicx}
\usepackage{booktabs}
\usepackage{graphicx}
\usepackage[normalem]{ulem}
\usepackage{multirow}
\usepackage{amsmath} 
\usepackage{bm}      
\usepackage{multirow}
\usepackage[table,xcdraw]{xcolor}
\usepackage{colortbl}
\usepackage[normalem]{ulem}
\useunder{\uline}{\ul}{}
\usepackage[ruled,vlined]{algorithm2e}
\usepackage{algpseudocode} 

\begin{document}
\title{SafeClick: Error-Tolerant Interactive Segmentation of Any Medical Volumes via Hierarchical Expert Consensus}
\titlerunning{Error-Tolerant Interactive Segmentation}

\author{Yifan Gao\inst{1,2,3}\textsuperscript{*}, Jiaxi Sheng\inst{1,3}\textsuperscript{*}, Wenbin Wu\inst{1,3}, Haoyue Li\inst{1,4}, Yaoxian Dong\inst{1,3}, Chaoyang Ge\inst{1,3}, Feng Yuan\inst{1,3}, Xin Gao\inst{2,3}\textsuperscript{\Envelope} }  
\authorrunning{Y. Gao et al.}
\institute{
    School of Biomedical Engineering (Suzhou), Division of Life Science and Medicine, University of Science and Technology of China, Hefei, China \and
    Shanghai Innovation Institute, Shanghai, China \and
    Suzhou Institute of Biomedical Engineering and Technology, Chinese Academy of Sciences, Suzhou, China \and
    College of Medicine and Biological Information Engineering, Northeastern University, Shenyang, China \\
}

\maketitle

\begingroup
\renewcommand\thefootnote{\textit{*}}
\footnotetext[1]{These authors contributed equally to this work.}
\endgroup

\begingroup
\renewcommand\thefootnote{\textsuperscript{\Envelope}}
\footnotetext[2]{Corresponding author}
\endgroup

\begin{abstract}
Foundation models for volumetric medical image segmentation have emerged as powerful tools in clinical workflows, enabling radiologists to delineate regions of interest through intuitive clicks. While these models demonstrate promising capabilities in segmenting previously unseen anatomical structures, their performance is strongly influenced by prompt quality. In clinical settings, radiologists often provide suboptimal prompts, which affects segmentation reliability and accuracy. To address this limitation, we present SafeClick, an error-tolerant interactive segmentation approach for medical volumes based on hierarchical expert consensus. SafeClick operates as a plug-and-play module compatible with foundation models including SAM 2 and MedSAM 2. The framework consists of two key components: a collaborative expert layer (CEL) that generates diverse feature representations through specialized transformer modules, and a consensus reasoning layer (CRL) that performs cross-referencing and adaptive integration of these features. This architecture transforms the segmentation process from a prompt-dependent operation to a robust framework capable of producing accurate results despite imperfect user inputs. Extensive experiments across 15 public datasets demonstrate that our plug-and-play approach consistently improves the performance of base foundation models, with particularly significant gains when working with imperfect prompts. The source code is available at https://github.com/yifangao112/SafeClick.

\end{abstract}
\section{Introduction}

Medical image segmentation plays a vital role in clinical applications, facilitating precise delineation of regions of interest within medical images \cite{gao2023anatomy,gao2024mba,gao2025wega}. Recent advancements in vision foundation models have enabled radiologists to segment these regions through intuitive prompts, such as points and bounding boxes. These interactive approaches significantly streamline diagnostic and therapeutic workflows \cite{wang2018interactive,luo2021mideepseg}.

The evolution of foundation models for volumetric medical image segmentation has expanded the capabilities of automated analysis beyond 2D images. Models like SAM 2 \cite{ravi2024sam} and MedSAM 2 \cite{zhu2024medical} have demonstrated promising performance across various medical imaging tasks \cite{chen2024sam2,xiong2024sam2,zhang2024unleashing,li2024adaptive}. However, a significant limitation emerges when these models receive imperfect prompts, leading to substantial performance degradation \cite{gao2024desam}. Studies have shown that even minor deviations in prompt placement can result in up to a 30\% decrease in segmentation performance \cite{huang2024segment}. Similarly, research indicates performance drops by nearly 20\% when point prompts are randomly selected instead of precisely placed \cite{wong2023scribbleprompt}. This sensitivity to prompt quality severely restricts the clinical applicability of such models, as radiologists often struggle to provide consistently perfect prompts in the fast-paced and complex clinical environment.

To address this challenge, we introduce SafeClick, an error-tolerant interactive segmentation approach for medical volumes based on hierarchical expert consensus. SafeClick transitions from reliance on a single prompt-driven process to a collaborative decision-making framework. Our approach consists of two key components: the collaborative expert layer (CEL) and the consensus reasoning layer (CRL). The CEL comprises three specialized transformer modules: one processing intermediate image features, another analyzing final image features independently of prompts, and a third integrating prompt information. The CRL then dynamically fuses these complementary perspectives through cross-referencing and feature aggregation, emphasizing the most reliable features when prompt quality varies. This hierarchical consensus mechanism enables SafeClick to maintain high performance across diverse prompt conditions, enhancing reliability in clinical settings.

\begin{figure}[htbp]
	\includegraphics[width=1.0\textwidth]{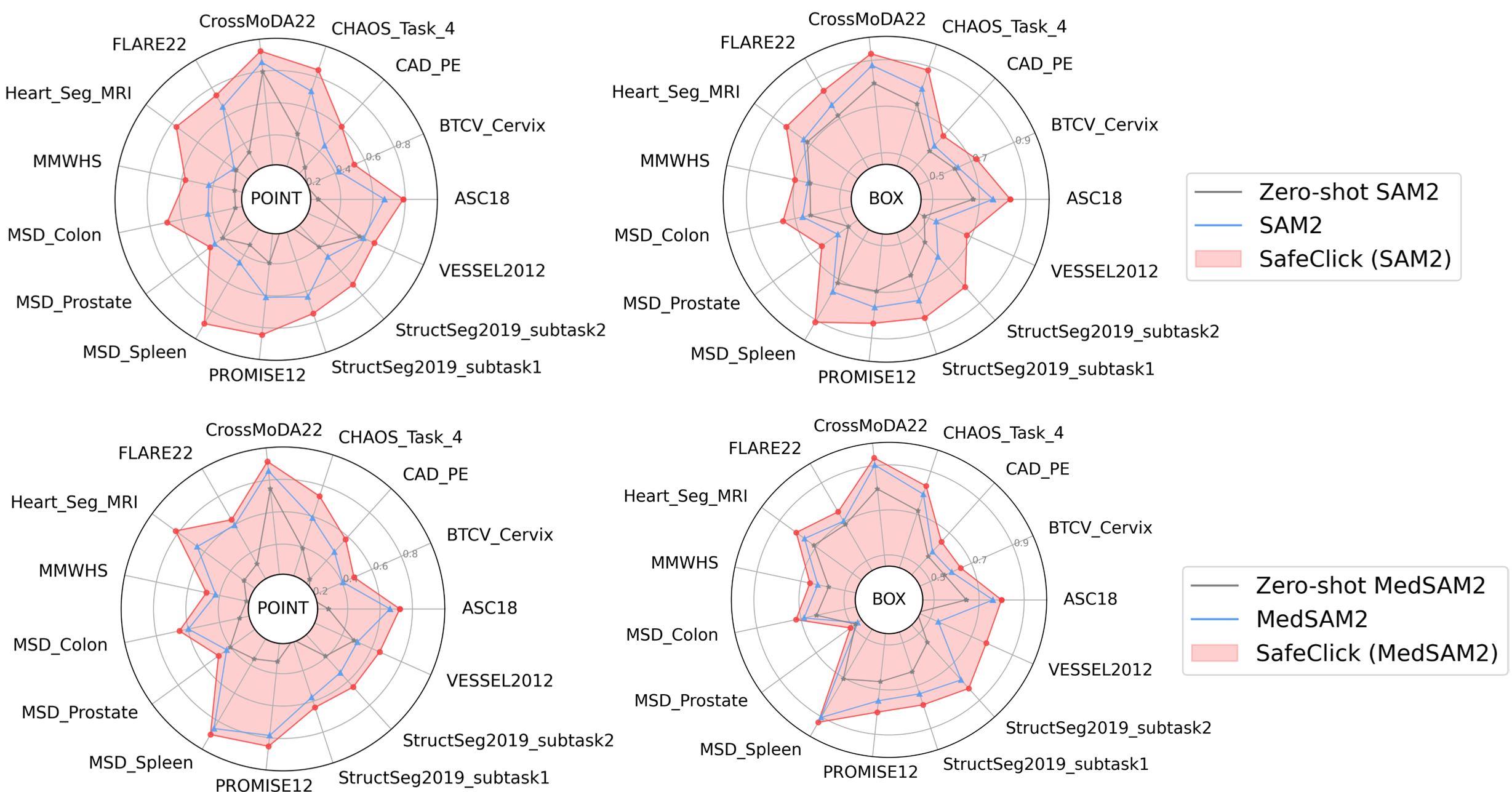}
	\centering
	\caption{Performance comparison between SafeClick and baseline foundation models across different datasets.} 
	\label{fig0}
\end{figure}

Extensive experiments across 15 public datasets with diverse anatomical structures demonstrate that SafeClick outperforms state-of-the-art foundation models, achieving superior performance with both ideal and imperfect prompts. As visualized in Fig. \ref{fig0}, SafeClick consistently outperforms baseline foundation models across multiple datasets, demonstrating its robust generalization capabilities across diverse anatomical structures. Our contributions are summarized as follows: (1) We present a plug-and-play module that enhances foundation models' resilience against imperfect prompts without requiring architectural modifications; (2) We introduce a hierarchical expert consensus mechanism that effectively balances prompt-dependent and image-intrinsic features; (3) We demonstrate consistent performance improvements across diverse anatomical structures and imaging modalities, establishing SafeClick's broad applicability in clinical scenarios.

\section{Methodology}
\begin{figure}[t!]
	\includegraphics[width=1.0\textwidth]{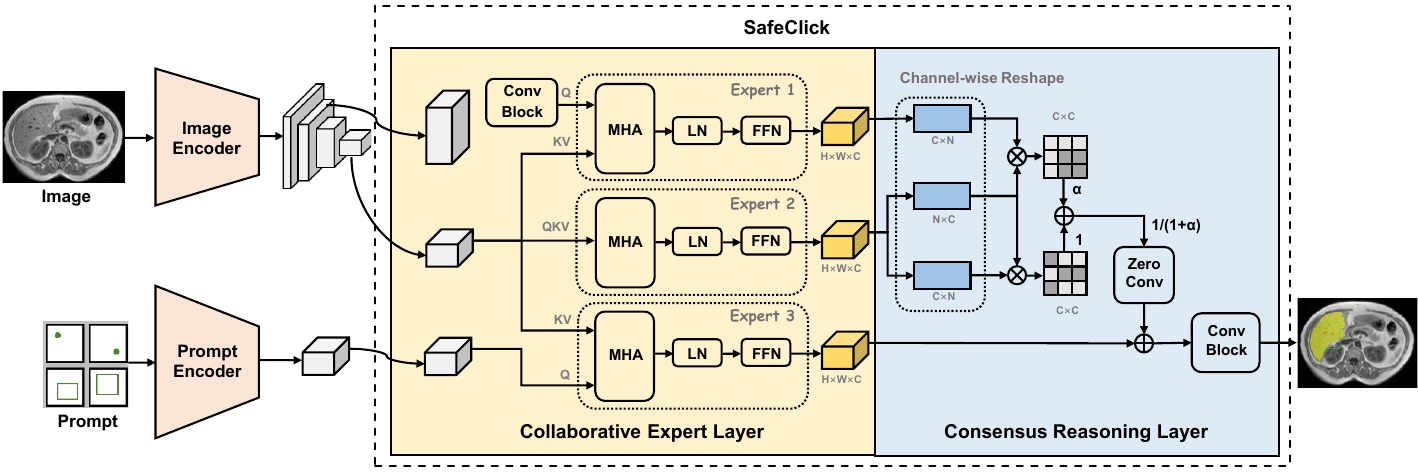}
	\centering
	\caption{Architecture of the proposed SafeClick. The diagram illustrates the two primary components: the collaborative expert layer (CEL) that generates diverse feature representations through specialized transformer modules, and the consensus reasoning layer (CRL) that performs cross-referencing and adaptive integration of these features. The framework operates as a plug-and-play module compatible with foundation models for medical volume segmentation.} 
	\label{fig1}
\end{figure}

Fig. \ref{fig1} presents the overall architecture of our proposed module. It adopts the same encoder as recent foundation models for medical volume segmentation, leveraging pre-trained weights for efficient feature extraction. The decoder of SafeClick is distinctively composed of two complementary components: a collaborative expert layer (CEL) and a consensus reasoning layer (CRL). The CEL is tasked with analyzing and refining features across different representational spaces, comprising three expert layers equipped with self-attention or cross-attention mechanisms. In parallel, the CRL integrates multi-level features through cross-referencing and feature aggregation, optimizing the final segmentation output.

\textbf{Collaborative Expert Layer: } The collaborative expert paradigm draws from the principle of divide-and-conquer \cite{jacobs1991adaptive}, breaking down complex segmentation tasks into manageable components that are addressed by specialized experts. Rather than relying solely on prompt-driven features, SafeClick distributes responsibility across multiple expert modules, each contributing complementary information to the segmentation process.

As shown in Fig. \ref{fig1}, the designed CEL within our framework is composed of three distinct transformer layers, each serving a specialized role in the process of feature analysis and refinement. This network includes two cross-attention transformer layers, designated as \(E_1\) and \(E_3\), alongside a self-attention transformer layer, referred to as \(E_2\). Notably, the architectural design of \(E_3\) mirrors the transformer layer found within the original mask decoder of foundation models.

Given inputs from the image encoder ($x_i \in \mathbb{R}^{H_p\times W_p \times C}$ and $x_f \in \mathbb{R}^{H\times W \times C}$) and prompt encoder ($x_p \in \mathbb{R}^{H\times W \times C}$), the CEL processes these representations through its specialized components. We select the $m/2$-th layer output from the $m$-layer image encoder as $x_i$ to capture multi-scale information.

The intermediate feature $x_i$ first undergoes a dimension transformation to match the spatial resolution of other features:
\begin{equation}
\hat{x}_i = \mathcal{F}_{transform}(x_i) \in \mathbb{R}^{H\times W \times C}
\end{equation}
where $\mathcal{F}_{transform}$ is a sequential operation including convolution, normalization, and activation. The first expert $E_1$ applies cross-attention between transformed intermediate and final image features:
\begin{equation}
\Tilde{x}_1 = \mathrm{MCA} \bigl( \text{LN} \bigl( \mathcal{T}_{cl}(\hat{x}_i) \bigr), \text{LN} \bigl( \mathcal{T}_{cl}(x_f) \bigr) \bigr) + \mathcal{T}_{cl}(\hat{x}_i)
\end{equation}
\begin{equation}
\quad x_1 = \mathcal{R} \bigl( \mathrm{MLP}(\mathrm{LN}(\Tilde{x}_1)) + \Tilde{x}_1 \bigr)
\end{equation}
where LN is layer normalization and MCA is multi-head cross-attention. $\mathcal{T}_{cl}(\cdot)$ is the channel last transformation, which transforms the spatial features with dimensions $H \times W \times C$ into a 2D embedding of $N \times C$ to capture spatial relationships in attention computation. $\mathcal{R}(\cdot)$ transform the 2D embedding back to the image dimensions. The second expert $E_2$ applies self-attention to final image features: 
\begin{equation}
\Tilde{x}_2 = \mathrm{MSA}\bigl(\text{LN} \bigl( \mathcal{T}_{cl}(x_f) \bigr)\bigr) + \mathcal{T}_{cl}({x}_f)
\end{equation} 
\begin{equation}
\quad x_2 = \mathcal{R} \bigl( \mathrm{MLP}(\mathrm{LN}(\Tilde{x}_2)) + \Tilde{x}_2 \bigr)
\end{equation} 
where MSA is multi-head self-attention, enabling prompt-independent analysis of image content.

The third expert $E_3$ processes prompt features $x_p$ with $x_f$ using cross-attention, producing output $x_3$. Through this architectural composition, our CEL not only enriches the feature space available for segmentation tasks but also introduces a level of adaptability and specificity that is essential for handling the variability inherent in medical volumes and user prompts.

\textbf{Consensus Reasoning Layer: } The Consensus Reasoning Layer (CRL) distills complementary information from the expert modules to handle imperfect user prompts in medical volume segmentation. Given outputs $x_1, x_2 \in \mathbb{R}^{H \times W \times C_1}$ from our expert layers and prompt-dependent features $x_3 \in \mathbb{R}^{H \times W \times C}$, the CRL performs cross-referencing between these representations.

We first transform the feature tensors into channel-first representation:
\begin{equation}
\Phi_1 = \mathcal{T}(x_1) \in \mathbb{R}^{C \times N}, \quad \Phi_2 = \mathcal{T}(x_2) \in \mathbb{R}^{C \times N}
\end{equation}
where $\mathcal{T}(\cdot)$ denotes reshaping operation and $N = H \times W$.

We compute and normalize cross-reference attention matrices through a contrastive mechanism:
\begin{equation}
\hat{\mathcal{A}}_{cross} = \max(\Phi_1 \Phi_2^\top) \cdot \mathbb{1} - \Phi_1 \Phi_2^\top, \quad \hat{\mathcal{A}}_{self} = \max(\Phi_2 \Phi_2^\top) \cdot \mathbb{1} - \Phi_2 \Phi_2^\top
\end{equation}
where $\mathbb{1}(\cdot)$ denotes the all-one matrix.

These attention matrices are combined through a learnable parameter $\alpha$ and applied to modulate features:
\begin{equation}
x'_2 = \mathcal{R}\left(\frac{\sigma(\hat{\mathcal{A}}_{self}) + \alpha \cdot \sigma(\hat{\mathcal{A}}_{cross})}{1 + \alpha} \Phi_2 + \Phi_2\right)
\end{equation}
where $\sigma(\cdot)$ is softmax operation and $\mathcal{R}(\cdot)$ reshapes back to spatial dimensions.

The prompt-guided features are integrated with this refined representation:
\begin{equation}
x_{enhanced} = \phi(\mathcal{N}(\mathcal{F}_{conv}(x'_2) + x_3))
\end{equation}
where $\mathcal{F}_{conv}$ is a convolutional operation with zero-initialized parameters, ensuring the network initially preserves the prompt-driven features while gradually learning optimal feature integration during training. $\mathcal{N}(\cdot)$ is layer normalization, and $\phi(\cdot)$ represents leaky ReLU activation.

This consensus building process enables our framework to balance multiple expert perspectives, maintaining high segmentation performance even with suboptimal user inputs. The output $x_{enhanced}$ is passed to upsampling layers for generating the final segmentation mask.

\section{Experiments and Results}

\textbf{Dataset and Preprocessing: } We evaluate SafeClick on 15 public 3D medical datasets spanning various anatomical regions including brain, abdomen, lung, cardiac, urology, gynecology, and vasculature structures, as shown in Table 1. These datasets collectively contain over 89,000 annotated regions of interest across different imaging modalities. For preprocessing, we normalize intensity values to the range [0, 1] and resample all volumes to isotropic spacing \cite{ye2023sa}. We simulate imperfect prompts by adding noise to perfect prompts: for point prompts, we apply random displacements from the center of mass ranging from 25\% to 100\% of the object radius; for bounding boxes, we scale the perfect box by factors ranging from 50\% to 150\%.

\begin{table}[htbp]
\centering
\caption{Overview of the 15 public datasets used for evaluation, organized by body region. The table shows the anatomical focus and number of regions of interest (ROIs) in each dataset, covering a diverse range of medical imaging applications.}
\label{tab:1}
\resizebox{\columnwidth}{!}{%
\begin{tabular}{ccc|ccc|ccc}
\toprule
Dataset      & Body       & ROIs & Dataset       & Body    & ROIs  & Dataset                & Body        & ROIs  \\ \midrule
ASC18 \cite{ASC18}        & Brain      & 8855 & FLARE22 \cite{FLARE22}       & Brain   & 23368 & MSD\_Spleen \cite{MSD}             & Abdomen     & 1008  \\
BTCV\_Cervix \cite{BTCV_Cervix}  & Gynecology & 4667 & Heart\_Seg\_MRI \cite{Heart_Seg_MRI} & Cardiac & 517   & PROMISE12 \cite{PROMISE12}              & Urology     & 776   \\
CAD\_PE \cite{CAD_PE}       & Lung       & 3102 & MMWHS \cite{MMWHS4}         & Brain   & 17684 & StructSeg2019\_subtask1 \cite{StructSeg2019} & Abdomen     & 3688  \\
CHAOS\_Task\_4 \cite{CHAOS_Task_4} & Abdomen    & 3513 & MSD\_Colon \cite{MSD}     & Abdomen & 1093  & StructSeg2019\_subtask2 \cite{StructSeg2019} & Brain       & 4781  \\
CrossMoDA22 \cite{CrossMoDA22}  & Brain      & 1478 & MSD\_Prostate \cite{MSD}  & Urology & 1466  & VESSEL2012 \cite{VESSEL2012}            & Vasculature & 13182 \\ \bottomrule
\end{tabular}%
}
\end{table}

\textbf{Implementation and Experiment Setting: } We implement SafeClick as a plug-and-play module compatible with SAM 2 and MedSAM 2 across all their model sizes (ViT-T, ViT-S, ViT-L). For each dataset, we implemented random partitioning into training, validation, and test sets in a 7:1:2 ratio. Our implementation uses PyTorch, with training performed on NVIDIA H100 GPUs. We train with AdamW optimizer, using a learning rate of 1e-4 with cosine annealing schedule, batch size of 8, and train for 20 epochs. For a fair comparison, following the same approach as fine-tuning SAM2 and MedSAM2, we froze the encoder weights of the base model and trained only the SafeClick module. As a lightweight module, SafeClick introduces only an 18\% increase in inference time compared to the baseline architectures. We evaluate performance using Dice similarity coefficient and compare results under both perfect and imperfect prompt conditions.

\begin{figure}
	\includegraphics[width=0.9\textwidth]{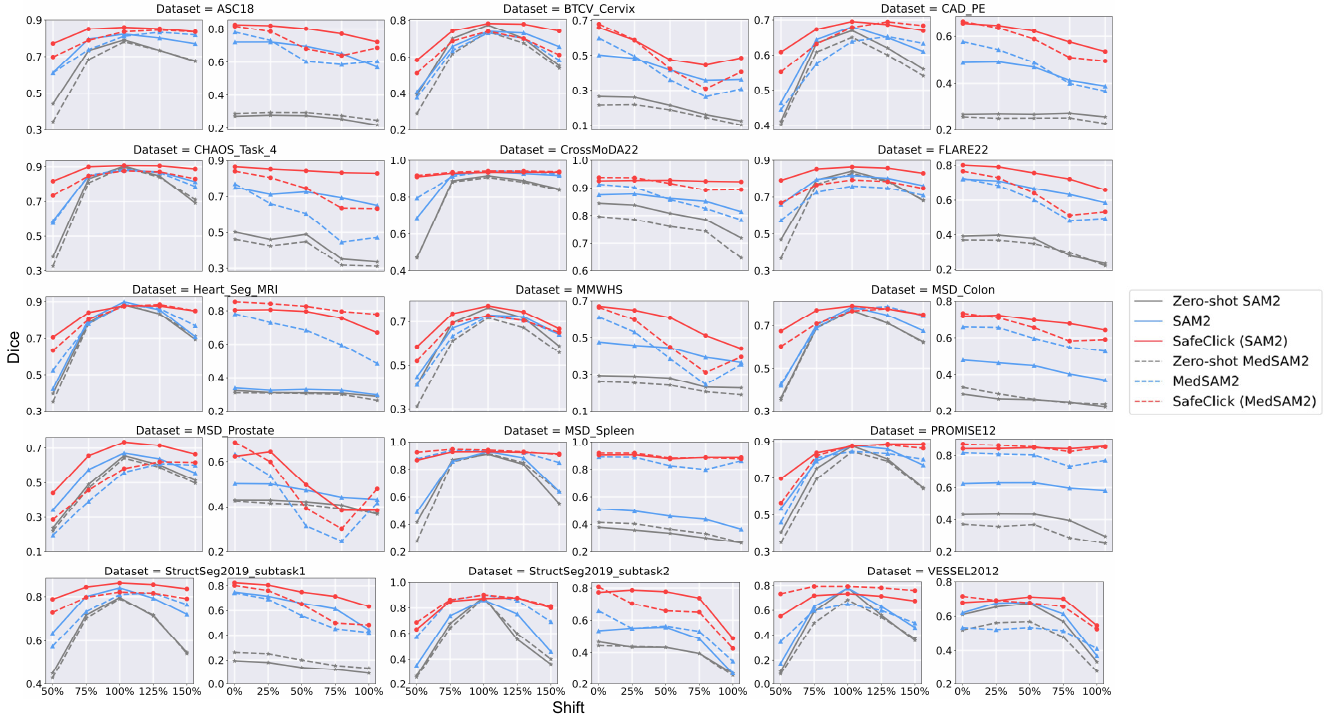}
	\centering
	\caption{Detailed results per dataset with box prompt (left) and point prompt (right).} 
	\label{fig3}
\end{figure}

\begin{figure}
	\includegraphics[width=0.7\textwidth]{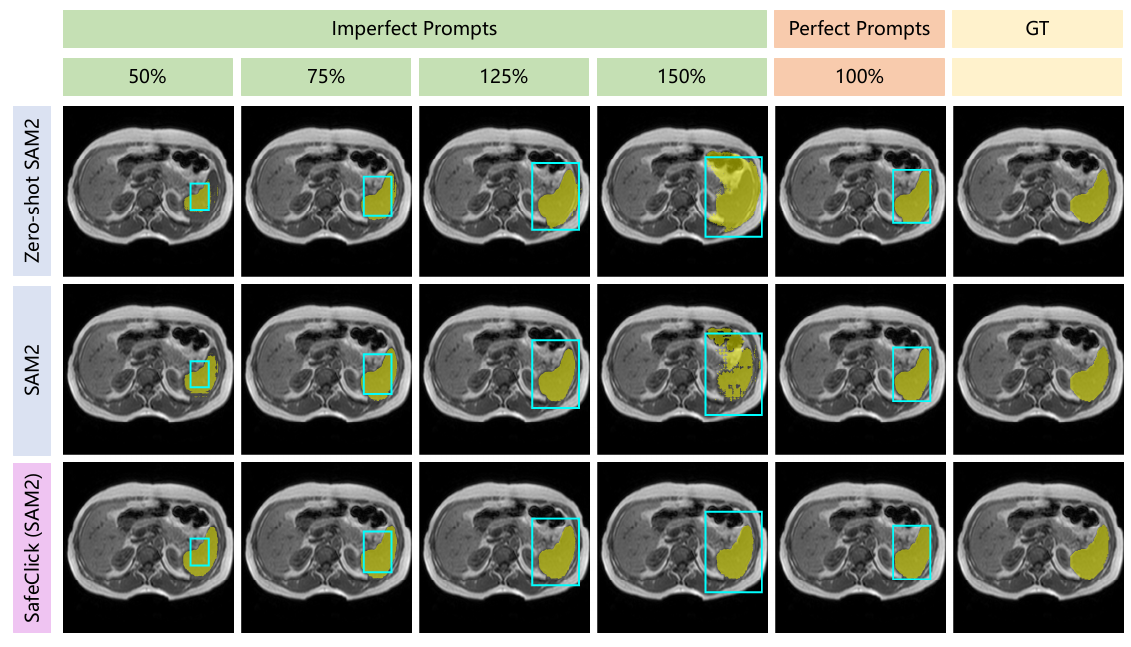}
	\centering
	\caption{Qualitative results in the CHAOS dataset under box prompts. The red star indicates the location of the point prompt. GT: ground truth.} 
	\label{fig2}
\end{figure}

\begin{table}[t!]
\centering
\caption{Performance comparison (Dice scores (\%)) of SafeClick against baseline models under perfect prompts (PP) and imperfect prompts (IP) of varying quality. These results represent the average performance across 15 datasets. For point prompts, percentages indicate displacement from ideal position; for bounding boxes, percentages indicate scale relative to perfect box. Bold and underlined values indicate best and second-best results.}
\label{tab:2}
\resizebox{0.9\columnwidth}{!}{%
\begin{tabular}{c|c|cccccccccccc}
\toprule
                                       &                               & \multicolumn{6}{c|}{Point}                                                                                                                                                                                                        & \multicolumn{6}{c}{Bbox}                                                                                                                                                                                 \\ \midrule
                                       &                               & \multicolumn{1}{c|}{PP}                          & \multicolumn{5}{c|}{IP}                                                                                                                                                       & \multicolumn{1}{c|}{PP}                          & \multicolumn{5}{c}{IP}                                                                                                                                \\
\multirow{-2}{*}{Methods}              & \multirow{-2}{*}{Model Types} & \multicolumn{1}{c|}{0\%}                         & 25\%                         & 50\%                         & 75\%                         & 100\%                        & \multicolumn{1}{c|}{Avg.}                         & \multicolumn{1}{c|}{100\%}                       & 50\%                         & 75\%                        & 125\%                       & 150\%                        & Avg.                        \\ \midrule
                                       & ViT-T                         & \multicolumn{1}{c|}{44.08}                       & 43.72                        & 42.69                        & 38.57                        & 33.79                        & \multicolumn{1}{c|}{39.69}                        & \multicolumn{1}{c|}{81.33}                       & 35.15                        & 71.90                       & 73.20                       & 60.68                        & 60.23                       \\
                                       & ViT-S                         & \multicolumn{1}{c|}{40.05}                       & 39.35                        & 38.05                        & 32.47                        & 25.63                        & \multicolumn{1}{c|}{33.88}                        & \multicolumn{1}{c|}{81.66}                       & 36.72                        & 71.76                       & 72.73                       & 57.42                        & 59.66                       \\
\multirow{-3}{*}{Zero-shot SAM 2}      & ViT-L                         & \multicolumn{1}{c|}{35.15}                       & 34.17                        & 33.75                        & 30.58                        & 25.65                        & \multicolumn{1}{c|}{31.04}                        & \multicolumn{1}{c|}{81.25}                       & 40.51                        & 73.16                       & 72.16                       & 59.59                        & 61.36                       \\ \midrule
                                       & ViT-T                         & \multicolumn{1}{c|}{56.06}                       & 55.85                        & 54.58                        & 51.10                        & 42.76                        & \multicolumn{1}{c|}{51.07}                        & \multicolumn{1}{c|}{81.53}                       & 40.52                        & 72.59                       & 76.08                       & 65.66                        & 63.71                       \\
                                       & ViT-S                         & \multicolumn{1}{c|}{60.14}                       & 59.38                        & 56.95                        & 52.59                        & 45.30                        & \multicolumn{1}{c|}{53.56}                        & \multicolumn{1}{c|}{81.00}                       & 50.73                        & 75.94                       & 77.33                       & 67.29                        & 67.82                       \\
\multirow{-3}{*}{SAM 2}                & ViT-L                         & \multicolumn{1}{c|}{61.68}                       & 61.00                        & 58.86                        & 54.44                        & 49.27                        & \multicolumn{1}{c|}{55.89}                        & \multicolumn{1}{c|}{82.36}                       & 53.52                        & 76.02                       & 79.72                       & 70.21                        & 69.87                       \\ \midrule
                                       & ViT-T                         & \multicolumn{1}{c|}{75.00}                       & 74.29                        & 71.11                        & 67.72                        & 62.95                        & \multicolumn{1}{c|}{69.02}                        & \multicolumn{1}{c|}{82.28}                       & 66.13                        & 79.08                       & 81.74                       & 78.75                        & 76.43                       \\
                                       & ViT-S                         & \multicolumn{1}{c|}{{\ul 78.06}}                 & {\ul 77.33}                  & {\ul 74.03}                  & {\ul 70.88}                  & \textbf{65.79}               & \multicolumn{1}{c|}{{\ul 72.01}}                  & \multicolumn{1}{c|}{{\ul 83.57}}                 & \textbf{71.22}               & \textbf{81.76}              & {\ul 82.93}                 & \textbf{80.57}               & \textbf{79.12}              \\
\multirow{-3}{*}{SafeClick (SAM 2)}    & ViT-L                         & \multicolumn{1}{c|}{\textbf{78.38}}              & \textbf{77.94}               & \textbf{74.66}               & \textbf{71.09}               & {\ul 65.31}                  & \multicolumn{1}{c|}{\textbf{72.25}}               & \multicolumn{1}{c|}{\textbf{83.88}}              & {\ul 71.06}                  & {\ul 81.63}                 & \textbf{82.95}              & {\ul 79.34}                  & {\ul 78.75}                 \\ \midrule
Improve                                &                               & \multicolumn{1}{c|}{{\color[HTML]{FE0000} 17.85}} & {\color[HTML]{FE0000} 17.78} & {\color[HTML]{FE0000} 16.47} & {\color[HTML]{FE0000} 17.19} & {\color[HTML]{FE0000} 18.91} & \multicolumn{1}{c|}{{\color[HTML]{FE0000} 17.59}} & \multicolumn{1}{c|}{{\color[HTML]{FE0000} 1.61}} & {\color[HTML]{FE0000} 21.21} & {\color[HTML]{FE0000} 5.97} & {\color[HTML]{FE0000} 4.83} & {\color[HTML]{FE0000} 11.83} & {\color[HTML]{FE0000} 10.96} \\ \midrule
                                       & ViT-T                         & \multicolumn{1}{c|}{43.84}                       & 42.91                        & 41.50                        & 36.60                        & 30.93                        & \multicolumn{1}{c|}{37.99}                        & \multicolumn{1}{c|}{80.19}                       & 34.11                        & 69.72                       & 73.29                       & 60.73                        & 59.46\\
                                       & ViT-S                         & \multicolumn{1}{c|}{40.29}                       & 39.69                        & 38.01                        & 33.99                        & 26.75                        & \multicolumn{1}{c|}{34.61}                        & \multicolumn{1}{c|}{78.84}                       & 31.06                        & 67.10                       & 70.88                       & 57.66                        & 56.68\\
\multirow{-3}{*}{Zero-shot MedSAM 2}   & ViT-L                         & \multicolumn{1}{c|}{30.37}                       & 29.91                        & 29.43                        & 25.82                        & 22.53                        & \multicolumn{1}{c|}{26.92}                        & \multicolumn{1}{c|}{79.34}                       & 31.60                        & 69.23                       & 72.15                       & 61.05                        & 58.51\\ \midrule
                                       & ViT-T                         & \multicolumn{1}{c|}{70.87}                       & 65.81                        & 56.65                        & 49.57                        & 49.61                        & \multicolumn{1}{c|}{55.41}                        & \multicolumn{1}{c|}{78.64}                       & 52.96                        & 72.40                       & 77.53                       & 71.55                        & 68.61\\
                                       & ViT-S                         & \multicolumn{1}{c|}{72.23}                       & 67.72                        & 61.42                        & 53.99                        & 53.14                        & \multicolumn{1}{c|}{59.07}                        & \multicolumn{1}{c|}{79.94}                       & 53.70                        & 73.53                       & 79.18                       & 74.00                        & 70.10\\
\multirow{-3}{*}{MedSAM 2}             & ViT-L                         & \multicolumn{1}{c|}{70.74}                       & 65.04                        & 57.55                        & 49.83                        & 49.79                        & \multicolumn{1}{c|}{55.55}                        & \multicolumn{1}{c|}{78.54}                       & 48.84                        & 71.41                       & 78.03                       & 70.76                        & 67.26\\ \midrule
                                       & ViT-T                         & \multicolumn{1}{c|}{77.75}                       & 73.52                        & 65.34                        & 57.54                        & 58.52                        & \multicolumn{1}{c|}{63.73}                        & \multicolumn{1}{c|}{80.62}                       & 64.48                        & 76.67                       & 80.21                       & 77.04                        & 74.60\\
                                       & ViT-S                         & \multicolumn{1}{c|}{\textbf{79.06}}              & \textbf{75.01}               & \textbf{68.22}               & {\ul 61.24}                  & \textbf{61.64}               & \multicolumn{1}{c|}{\textbf{66.53}}               & \multicolumn{1}{c|}{\textbf{81.29}}              & {\ul 65.02}                  & {\ul 76.98}                 & \textbf{81.09}              & \textbf{78.55}               & \textbf{75.41}\\
\multirow{-3}{*}{SafeClick (MedSAM 2)} & ViT-L                         & \multicolumn{1}{c|}{{\ul 78.41}}                 & {\ul 74.79}                  & {\ul 67.23}                  & \textbf{61.36}               & {\ul 60.91}                  & \multicolumn{1}{c|}{{\ul 66.07}}                  & \multicolumn{1}{c|}{{\ul 81.06}}                 & \textbf{65.67}               & \textbf{77.06}              & {\ul 80.78}                 & {\ul 77.32}                  & {\ul 75.21}\\ \midrule
Improve                                &                               & \multicolumn{1}{c|}{{\color[HTML]{FF0000} 7.13}} & {\color[HTML]{FF0000} 8.25}  & {\color[HTML]{FF0000} 8.39}  & {\color[HTML]{FF0000} 8.92}  & {\color[HTML]{FF0000} 9.51}  & \multicolumn{1}{c|}{{\color[HTML]{FF0000} 8.77}}  & \multicolumn{1}{c|}{{\color[HTML]{FF0000} 1.95}} & {\color[HTML]{FF0000} 13.22} & {\color[HTML]{FF0000} 4.46} & {\color[HTML]{FF0000} 2.45} & {\color[HTML]{FF0000} 5.53}  & {\color[HTML]{FF0000} 6.42}\\ \bottomrule
\end{tabular}%
}
\end{table}

\textbf{Results: } Table \ref{tab:2} and Fig. \ref{fig3} report the quantitative comparison of various methods across 15 public datasets, with Fig. \ref{fig2} offering visual comparisons of segmentation results. SafeClick consistently improves the performance of both SAM 2 and MedSAM 2 across all prompt conditions. For SAM 2, our method achieves average improvements of 17.59\% and 10.96\% in Dice score for imperfect point and bounding box prompts, respectively. For MedSAM 2, the average improvements for imperfect prompts are 8.77\% and 6.42\%. Notably, performance gains are most significant under challenging conditions with highly imperfect prompts, where baseline models experience substantial degradation. For instance, with 50\% displaced box prompts, SafeClick improves SAM 2's performance by 21.21\%. The results demonstrate that our method effectively maintains segmentation quality even when prompt quality deteriorates, confirming its value in practical clinical scenarios where perfect user interaction cannot be guaranteed.

\textbf{Ablation Study: } We conduct ablation studies on the ASC18 dataset to evaluate the contribution of each component in our SafeClick framework, with results presented in Table 3. Removing the cross-attention transformer layer ($E_1$) results in performance drops of 2.93\% and 1.44\% for point and bounding box prompts under imperfect conditions, indicating its importance in cross-referencing multi-level features. When the self-attention transformer layer ($E_2$ is removed, performance decreases by 5.15\% and 2.98\%, suggesting this component's significant role in providing prompt-independent analysis of image content. Specially, the hybrid-attention transformer layer ($E_3$) refers to the Transformer layer of the original mask decoder in SAM2 to integrate the prompt with image features. Removing $E_3$ would prevent the model from processing the prompt and thus change the model's prompt interaction mode. As this paper focuses on researching prompts of different qualities, the $E_3$ expert layer is retained. The absence of the CRL leads to reductions of 4.53\% and 2.30\%, confirming its effectiveness in adaptively integrating expert outputs. These results demonstrate that each component contributes substantially to SafeClick's overall performance, with their collaborative interaction yielding optimal results. 

\begin{table}[htbp]
\centering
\caption{Ablation studies of the various components of Safeclick on ASC18 dataset.}
\label{tab:3}
\resizebox{\columnwidth}{!}{%
\begin{tabular}{c|cccccc|cccccc}
\toprule
Type &                        & Baseline & w/o \(E_1\) & w/o \(E_2\) & w/o CRL     & SafeClick      &                       & Baseline & w/o \(E_1\) & w/o \(E_2\) & w/o CRL     & SafeClick            \\ \midrule
PP            & \multirow{2}{*}{Point} & 77.19    & 80.54       & 78.31       & {\ul 80.99} & \textbf{83.46} & \multirow{2}{*}{Bbox} & 83.05    & 84.20       & 83.46       & {\ul 85.09} & \textbf{86.46}       \\
IP Avg.       &                        & 72.06    & {\ul 76.98} & 74.76       & 75.38       & \textbf{79.91} &                       & 77.89    & {\ul 81.55}       & 80.01       & 80.69 & \textbf{82.99} \\ \bottomrule
\end{tabular}%
}
\end{table}

\section{Discussion and Conclusion}
We presented SafeClick, a plug-and-play error-tolerant module for interactive medical volume segmentation that significantly enhances foundation models' resilience against imperfect prompts. SafeClick's hierarchical expert consensus mechanism can be viewed from an ensemble learning perspective, where multiple specialized experts analyze the input data through complementary pathways, reducing the variance of predictions when faced with noisy inputs. This aligns with classical mixture-of-experts model where combining diverse estimators leads to improved performance. 

Our extensive evaluation across 15 datasets shows interesting patterns regarding different prompt types. Point prompts, which provide minimal spatial guidance, benefit from consistent improvements regardless of prompt quality, as SafeClick's collaborative experts compensate for the limited information by leveraging image features. Bounding box prompts, while generally more robust in baseline models, show particularly substantial gains under imperfect conditions (up to 21.21\% improvement for severely distorted boxes), which is especially valuable in clinical settings where precise box placement is challenging during time-constrained examinations.

A notable strength of SafeClick is its universal compatibility with foundation models built upon SAM2 architecture without requiring major architectural modifications, enabling straightforward integration into existing clinical workflows across diverse imaging modalities. The module adds minimal computational overhead (approximately 18\% additional time) while delivering substantial performance gains across all tested anatomical regions from brain to vasculature structures. While SafeClick significantly improves robustness, extremely poor prompts that completely miss the target region remain challenging. Future work could explore integrating SafeClick with automatic prompt correction mechanisms and extending its application to more diverse medical imaging types.

\bibliography{mybibliography}
\end{document}